# Quantum Behaviors and Dynamic Horizons[a]

James Lindesay

*Computational Physics Lab, Howard University, Washington, D.C.*

**Abstract.** Geometries with horizons offer insights into relationships between general relativity and quantum physics. Quantum mechanics constrains relationships between kinematic parameters and the coordinates describing the dynamics. Example quantum behaviors on space-times with dynamic horizons will be demonstrated, with an emphasis on examining co-gravitating quantum systems. Finally, the large scale causal structure of a multi-fluid cosmology that can describe dynamic coherent aspects of the universe as a whole will be presented.

## INTRODUCTION AND OUTLINE

An outstanding challenge of modern physics involves the consistent incorporation of quantum phenomena in gravitating environments. Attempts to develop microscopic gravitation using purely geometric considerations typically encounter the complications of describing locally linear quantum processes in terms of non-linear curvature effects. In order to allow the known phenomenology to help guide the discussion, the present approach will steadfastly cling to the few experimental results that involve both Newton's constant $G_N$ and Planck's constant $\hbar$ in coherent interrelation[1]. The commonly observed disentanglements of quantum processes in gravitating laboratories demonstrate no evidence of anomalous gravitational behaviors. Macroscopically coherent systems, like superfluids, gravitate consistent with the equivalence principle, apparently independent of the thermal fraction of the components. These observations should offer at least some clues as to the fundamentals of quantum geometrodynamics.

This presentation will focus on examining the quantum behaviors of gravitating systems in a macroscopically generated space-time. The approach has been to first develop and explore a dynamic geometry upon which quantum coherent processes can be explored in a straightforward manner[2,3]. The dynamic black hole that will serve as the space-time background has a dynamic horizon, as well as other features of interest, which provide surfaces near which quantum phenomena cannot be neglected. The behaviors of the well-understood Klein-Gordon scalar field on that dynamic background will next be explored and displayed. One might then ask if a superposition of such Klein-Gordon fields might not be able to self-gravitate the

---

dynamic geometry, and insights will be drawn from examination of such a possibility. Those insights guide one to construct non-interacting scalar fields that can co-gravitate to consistently generate their dynamic background. One discovers that there are numerous generic solutions that allow macroscopic co-gravitation, independent of the internal microscopic behaviors of the disentangled constituents[4].

The author initially began the exploration of this dynamic black hole geometry in order to gain insights into dynamic cosmological models previously examined[5]. Armed with the insights gained from the examination of the dynamic horizon associated with an excreting black hole, one then has better tools for dealing with the problems of dark energy in a dynamic cosmology. The global structure of a multi-fluid dynamic cosmology is finally presented as a prelude to the exploration of quantum behaviors in a dynamic cosmology. A brief outline of the presentation is given below:

> I. Entanglement and Geometry
> II. Gravitation of Quantum Systems
> III. Macroscopic Co-Gravitation
> IV. Global Structure of a Multi-fluid cosmology
> V. Conclusions and Discussion

# ENTANGLEMENT AND GEOMETRY

## Experimental Evidence

Experiments by Overhauser, et.al.[1] in the early to mid 1970's have demonstrated gravitation of coherent systems. In a series of experiments, neutrons were shown to maintain their coherence through double slit or Bragg scattering interference while gravitating. A diagram of the double slit apparatus is shown in Figure 1.

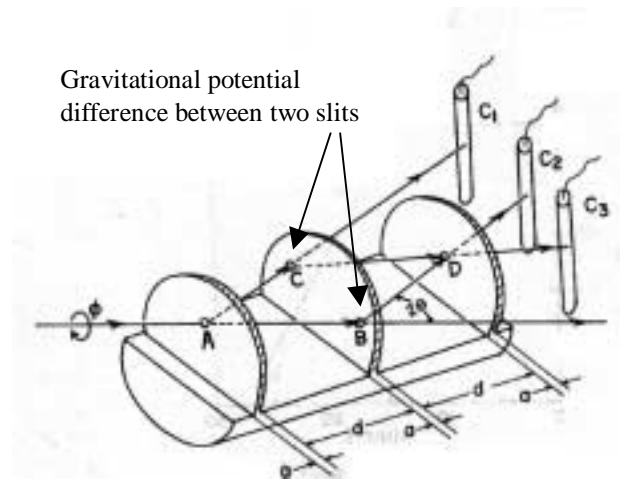

**FIGURE 1.** Overhauser's apparatus.

In the experiment, neutrons are collimated through A, coherently interfere through the two slits B and C, and are detected at detectors C1, C2, and C3. The entire apparatus can be rotated around the axis AB, resulting in gravitational phase differences for varying angles φ. The interference pattern associated with the count difference between detectors C2 and C3, as a function of the rotation angle, is demonstrated in Figure 2.

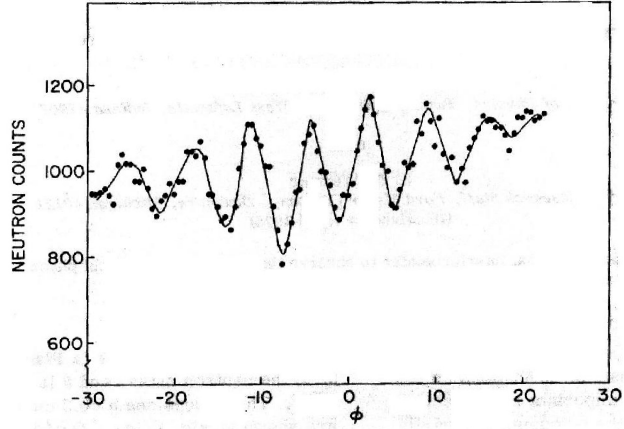

**FIGURE 2.** Neutron interference pattern, count difference $I_2$-$I_3$.

This result means that (at least for weak, slowly moving sources) gravitating systems maintain their quantum behaviors. The experiments were also confirmations of the principle of equivalence, i.e., that the motion of the observer does not break the coherence of an inertial system. In this case, the inertial neutron maintains coherence despite the nearly uniform acceleration of the laboratory apparatus. For the interested reader, a metric form for an observer with uniform acceleration has been developed in reference 8. An immediate corollary of this result was that the coherent neutrons in the Earth's gravitational field did not exchange "quanta of force" that would have localized those neutrons at either slit, breaking their coherence prior to detection at the detectors. More concisely stated for present purposes, gravitating quantum systems maintain their coherence properties.

One might consider other experiments that could examine the coherence properties of gravitating systems. For instance, superfluid helium behaves as a macroscopic quantum coherent system below its lambda transition temperature. When the normal fluid is rotated while it is cooled below the lambda temperature, the superfluid component maintains persistent angular momentum once the rotation of the vessel ceases. This persistent superflow is in the form of an array of vortices[9] of quantized circulation. $\oint \vec{v}_s \cdot d\vec{l} = \frac{2\pi\hbar}{m_{He}}$. Each vortex carries a fundamental quantum of angular momentum, and the array (or single vortex) can be stabilized and imaged. One expects that vortices oriented skewed to the vertical should precess in the Earth's gravitational field, as generically demonstrated in Figure 3.

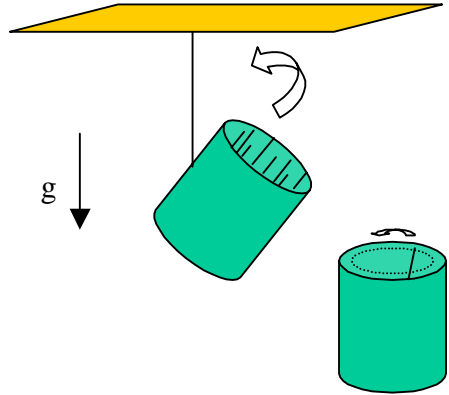

**FIGURE 3.** Expected quantized vortex precession.

The upper diagram suggests that the collection of previously imaged quantized vortices should combine to generate a measurable precession. Even single vortex states have been experimentally demonstrated, and whether their precession can be imaged as suggested in the lower diagram, or directly measured as in the upper depends upon the sensitivity of the experimental arrangement. Since a calculation of the precession frequency involves both Planck's constant and Newton's gravitational constant, such experiments give additional data on the coherence properties of gravitating systems. Any gravitationally induced precessions of the distributive motions that maintain coherent relational aspects in space-time make a strong statement about the fundamental nature of the gravitating behaviors of the disparate constituents of such a macroscopic quantum system.

## Quantum Measurement

Canonically conjugate physical parameters cannot be simultaneously measured by any single experiment on a quantum system. This places fundamental limits upon the extent to which properties of a quantum system can be observationally determined. Although one might attempt to perform many accurate experiments to arbitrarily increase the precision of knowledge about a system, the so called triangle inequality can be used to place lower limits upon the root mean squared (RMS) deviations of a large set of measurements from average values of conjugate parameters. This fundamental limit placed on quantum measurements is due to non-commutivity of operations, resulting in the uncertainty principles:

$$\begin{aligned} \Delta X\, \Delta P &\geq \frac{\hbar}{2} \\ \Delta t\, \Delta E &\geq \frac{\hbar}{2} \end{aligned}. \qquad (1)$$

In particular, the energy-time relation indicates that the quantum generation of a static (fixed energy) geometry is somewhat problematic. It is for this reason that a dynamic geometry will be explored in this presentation.

## *Spatial Coherence*

Quantum systems directly exhibit spatial coherence. This can be examined by exploring the space-like correlations associated with measurements of quantum-entangled events in the space-time. For instance, a straightforward calculation demonstrates that the vacuum expectation value of the symmetric sum of a field operator at different points is non-vanishing for space-like related events:

$$<vac|\phi(x)\phi(y)+\phi(y)\phi(x)|vac> = \frac{1}{4\pi^2 s^2}.$$
$$s^2 = |\vec{x}-\vec{y}|^2 - (x^0-y^0)^2 \qquad (2)$$

Since the vacuum expectation value of the field by itself is expected to vanish, this requires space-like correlations in measurements of the field at equal times $x^0 = y^0$:

$$\frac{1}{4\pi^2|\vec{x}-\vec{y}|^2} = <vac|\phi(x)\phi(y)+\phi(y)\phi(x)|vac>$$
$$\neq 2<vac|\phi(x)|vac><vac|\phi(y)|vac> = 0 \qquad (3)$$

Were any measurements independent, the two sides of the equation should be identical. However, the commutator of the (boson) field does vanish for space-like separations, which prevents a measurement at y from changing the probability distribution at x. Therefore, this microscopic causality forbids faster-than-light communications, but incorporates space-like coherence.

## GRAVITATION OF QUANTUM SYSTEMS

The behaviors of gravitating quantum systems are the primary topic of this section. As has been previously discussed, quantum systems exhibit spatial coherence. For this reason, the author has developed geometries that can serve as backgrounds for spatially coherent phenomena[2,3,5]. In order to gain insights into quantum phenomena in familiar gravitating environments, the author has chosen to begin by exploring quantum behaviors on a dynamic spherically symmetric geometry.

### Excreting Black Hole

The metric of the dynamic spherically symmetric space-time will be taken to have the form

$$ds^2 = -\left(1-\frac{R_M(ct)}{r}\right)c^2 dt^2 + 2\sqrt{\frac{R_M(ct)}{r}} cdt\, dr + dr^2 + r^2 d\theta^2 + r^2 \sin^2\theta\, d\varphi^2 . \quad (4)$$

This metric exhibits radial dynamics as indicated by the non-orthogonal temporal-radial component of the metric, as well as the temporal dependency of the parameter $R_M$, which has dimensions of length. The *radial mass scale* is defined as a dynamic form of the Schwarzschild radius $R_M(ct) \equiv 2G_N M(ct)/c^2$, where the gravitational source mass is dependent on this temporal coordinate. It should be noted that the coordinate $t$ is *not* the same as the time of a Schwarzschild observer, but it *is* the time measured by an asymptotic $(r \to \infty)$ observer in this geometry. The geometry changes coherently as the radial mass scale changes with coordinate $t$. Fixed $t$ surfaces are seen to remain space-like for all radial coordinates $r$. However, fixed radial surfaces are seen to transition from having time-like to space-like signatures at the radial mass scale $R_M$. The radial mass scale represents a surface for which outgoing photons would be momentarily stationary in the radial coordinate. However, for this geometry, the radial mass scale is not identical to the horizon (which itself *is* an outgoing light-like surface), but rather, for an excreting geometry $\dot{R}_M < 0$, lies outside of the horizon.

Next, the global causal structure of an evaporating black hole satisfying the metric form given in Eq. (4) will be examined. Conformal coordinates have been developed for the case of steady excretion[2]. Various useful properties of the resulting Penrose diagram can then be directly examined.

## *Dynamic Coordinate Grid*

The Penrose diagram of the radially dynamic excreting black hole is given in Figure 4.

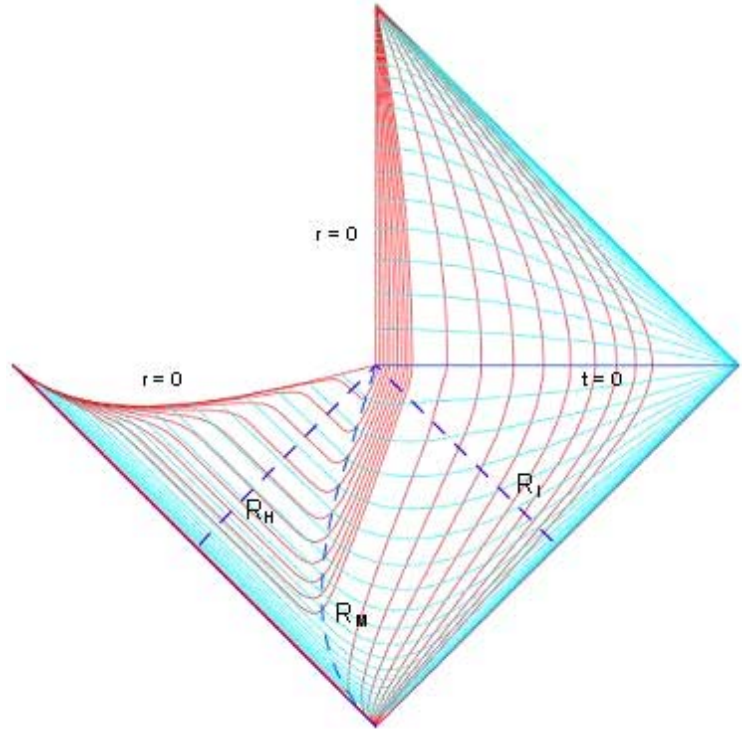

**FIGURE 4.** Penrose diagram of a spatially coherent black hole that excretes at a fixed rate.

Any given point *(r,ct)* on the Penrose diagram represents a sphere with area $4\pi r^2$ at time *t*. The curves that run horizontally on the right represent fixed time volumes, graded in units of the given scale. Curves that run vertically on the right represent fixed spherical areas with radial coordinate *r*, graded initially in tenths, then units of the given scale. The surface $R_H$ is the horizon of the black hole, and $R_M$ is the radial mass scale. The singularity *r=0* is the curve that bounds the left-hand portion of the diagram from above. The upper right-hand portion of the diagram represents space-time after the black hole has completely evaporated away (after *t=0*). The surface $R_I$ represents the last surface of communication with the singularity prior to its final evaporation. The reader is invited to more fully explore the details of this Penrose diagram in reference 2. The Penrose diagram will serve as a convenient medium upon which to display the quantum densities associated with the fields which will be explored in what follows.

## Klein-Gordon Fields On A Dynamic Black Hole Background

One of the most well understood quantum systems is the Klein-Gordon scalar field. The field equation of this scalar results from directly replacing the forms for the energy and momentum in the square of the non-interacting relativistic energy-momentum dispersion relation $E^2=p^2c^2+m^2c^4$ with their corresponding quantum operators:

$$\eta^{\mu\nu}\partial_\mu\partial_\nu\chi - \left(\frac{mc}{\hbar}\right)^2 \chi = \left(\nabla^2 - \frac{1}{c^2}\frac{\partial^2}{\partial t^2}\right)\chi - \left(\frac{mc}{\hbar}\right)^2 \chi = 0. \quad (5)$$

The behavior of a gravitating Klein-Gordon field can be directly examined by replacing the Minkowski metric in the Lagrangian that generates Eq. (5) with the metric appropriate to the background space-time[10]. If one examines the massless form of this equation in the metric space Eq. (4), after performing an angular momentum decomposition

$$\chi_{\ell m}(ct, r, \vartheta, \varphi) \equiv \frac{\psi_\ell(ct, r)}{r} Y_{\ell m}(\vartheta, \varphi). \quad (6)$$

one obtains dynamical equations for the massless Klein-Gordon field gravitating in the excreting black hole background:

$$-\frac{\partial^2 \psi_\ell}{(\partial ct)^2} + \frac{\partial}{\partial ct}\left[\sqrt{\frac{R_M}{r}}\left(\frac{\partial \psi_\ell}{\partial r}\right)\right] + \frac{\partial}{\partial r}\left[\sqrt{\frac{R_M}{r}}\left(\frac{\partial \psi_\ell}{\partial ct}\right)\right] + \frac{\partial}{\partial r}\left[\left(1 - \frac{R_M}{r}\right)\left(\frac{\partial \psi_\ell}{\partial r}\right)\right] +$$
$$- \left[\frac{\ell(\ell+1) + \frac{R_M}{r}}{r^2} + \frac{1}{r}\frac{\partial}{\partial ct}\sqrt{\frac{R_M}{r}}\right]\psi_\ell = 0. \quad (7)$$

It will be quite convenient to define a dimensionless scale associated with this space-time that will appear quite often in this presentation. The ratio of the temporally dependent radial mass scale to the radial coordinate will be defined as the parameter $\zeta \equiv \frac{R_M(ct)}{r}$. Assuming that the function $\psi$ depends on $ct$ and $r$ only through $\zeta$ allows Eq. (7) to be rewritten in terms of the single parameter $\zeta$:

$$\left(-\zeta + \zeta^{3/2} + \dot{R}_M\right)\left(\zeta + \zeta^{3/2} + \dot{R}_M\right)\frac{d^2\psi_\ell}{d\zeta^2} + 2\zeta^2\left(\zeta(-2 + 3\zeta) + 3\sqrt{\zeta}\dot{R}_M\right)\frac{d\psi_\ell}{d\zeta} +$$
$$\left(2\ell(\ell+1)\zeta^2 + 2\zeta^3 + 2m^2 R_M^2 + \zeta^{3/2}\dot{R}_M\right)\psi_\ell = 0. \quad (8)$$

Solutions to this differential equation in a single variable can be developed using straightforward methods. The probability density for a particular solution that has a very small likelihood of detecting the field on the surface $\zeta:=0$ (i.e., either asymptotically, or at t=0) has been plotted in Figure 5.

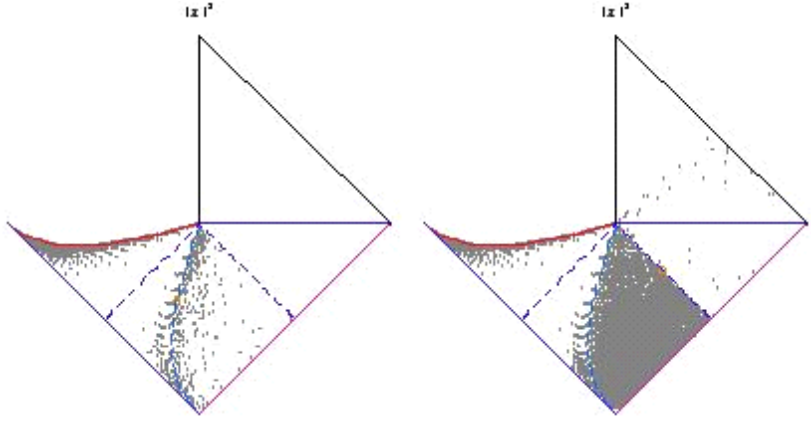

**FIGURE 5.** Penrose density plots for a gravitating massless Klein-Gordon scalar.

In these diagrams, the computer is instructed to place a gray pixel at a given location with coordinates *(ct,r)* if the magnitude squared of the gravitating Klein-Gordon solution $|\chi|^2$ has a value equal to or greater than its value at a chosen normalization point, indicated by the center of the small circles on either diagram. Both diagrams represent the same solution; only the normalization point has been altered. The diagram on the left has been normalized relative to a value on the radial mass scale, demonstrating that most of the density is near and within the horizon. The diagram on the right indicates the normalization scale at the ingoing horizon $R_I$, which means that most interior points will saturate. This diagram is included to demonstrate that there remains a small, coherent component to the Klein-Gordon field below the light-like surface communicating the end of evaporation, yet after the conclusion of the evaporation of the black hole. There is no field present above the light-like surface communicating the end of evaporation, as expected from causality arguments[6]. The energy density of the Klein-Gordon scalar can be calculated by examining the behavior of the action under metric variations $\delta W = \frac{1}{2}\int d^4x \sqrt{-g}\; T^{\mu\nu} \delta g_{\mu\nu}$. A straightforward calculation yields

$$T^{00} = \frac{1}{8\pi}\left[ |\chi'|^2 + |\dot\chi|^2 + \left(\frac{mc}{\hbar}\right)^2 |\chi|^2 + \zeta|\chi|^2 - 2\sqrt{\zeta}\,\Re e(\dot\chi^* \chi') \right]. \quad (9)$$

Considerable insight into the features required of a self-gravitating system can be gained by examining the functional form of the energy-momentum tensor of the Klein-Gordon scalar to that required to satisfy Einstein's equation for this geometry. In particular, the trace of Einstein's equation relates the Ricci scalar of the geometry **R** to the trace of the energy-momentum tensor. The functional forms of each of these densities are plotted in Figure 6.

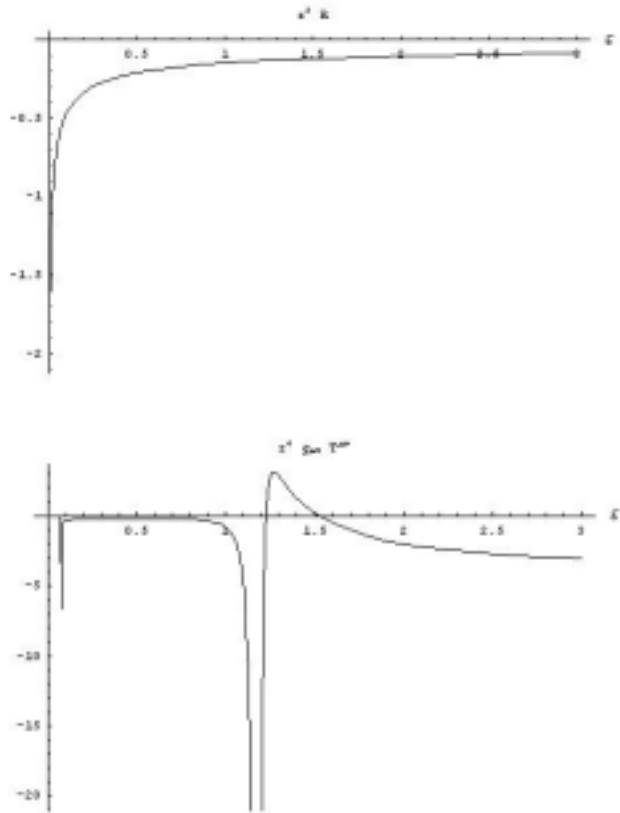

**FIGURE 6.** Ricci scalar of dynamic black hole (top) and trace of Klein Gordon tensor $T^\beta_\beta$ (bottom).

The functional form plotted in the top diagram is $r^2\,\mathbf{R}$ vs $\zeta$, while the bottom diagram represents $r^4\,T^\beta_\beta$ vs $\zeta$. Clearly, the differing dependencies on the radial coordinate (due to Planck length scales from Newton's gravitational constant, and dimensional derivatives in the field equations) make it problematic to combine gravitating Klein-Gordon fields in a manner to generate the background geometry. The radial dependency of the Klein-Gordon energy-momentum is a result of the quadratic form of derivatives in the Klein-Gordon Lagrangian generating the field equations. In order to construct co-gravitating scalar fields that generate the background geometry through Einstein's equation, these direct dimensional considerations must be met by the energy-momentum tensor of the constituent scalar fields. Both dimensional concerns, as well as methods to combine contributory quanta consistent with geometrodynamics, will be addressed in the next sections.

## Scattering Theory and Gravitation

Rather than attempt to develop a single microscopic field to self-gravitate in this geometry, the approach will be to develop disentangled fields that can co-gravitate consistently in the co-generated space-time background of the metric Eq. (4). This means that there is a need to intermingle the classical ideas in general relativity with quantum phenomenology. A viable formulation must incorporate the following fundamental tenets of the persistent quantum phenomena:

- Probability conservation / unitarity, expressed in the form $|\psi_{out}\rangle = S|\psi_{in}\rangle$, $\langle\psi_{out}|\psi_{out}\rangle = \langle\psi_{in}|\psi_{in}\rangle$, where S represents the scattering matrix;
- Lorentz covariance / the principle of equivalence; including an expectation of proper non-relativistic correspondence;
- Cluster decomposability / classical disentanglement; including an expectation that the kinematics of clusters parametrically add.

The implementation of these properties in flat space-time will next be reviewed.

## *Cluster Decomposability and Disentangled Scattering Theory*

Cluster decomposability describes how the various potentially interacting clusters in a composite system can be described in a disentangled manner. Without disentanglement, one cannot discuss the phenomena of classical physics in an meaningful way. For instance, one does not expect the behaviors of a few atoms on the moon to substantially affect a quantum scattering experiment here on Earth. Because of the non-linear energy-momentum dispersion relation associated with Lorentz covariance, the incorporation of off-shell quantum dynamics makes the analytic expression of cluster decomposability non-trivial. To more technically examine these concepts, examine Figure 7.

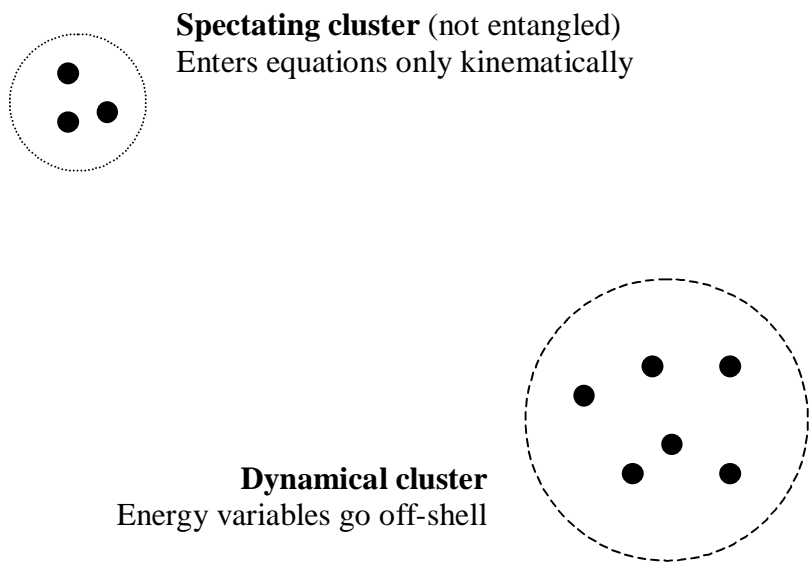

**FIGURE 7.** Clusters in quantum dynamics.

In the diagram, the dynamical cluster represents a coherent quantum system undergoing interactions, which are described by parameters that go *off-shell* during the intermediate scatterings. No constituent of the spectating cluster is entangled with any of the constituent reactants in the dynamical cluster during the process of interest. Therefore, the spectating cluster is classically disentangled from the quantum dynamics in the dynamical cluster.

For non-relativistic dynamics, Faddeev[11] successfully developed unitary cluster decomposable quantum scattering theory by defining channels based upon the possible dynamical clusters during a multi-particle scattering process. Utilizing Faddeev's channel decomposition, a relativistic version was later developed by the author and others[12,13]. The key elements of the relativistic solution are as follows:

- Lorentz frame velocity conservation should be used during intermediate dynamics rather than momentum conservation. If one examines the algebra of the Poincare' group, the boost, momentum, and energy generators satisfy the relation $[K_j, P_k] = i\,\delta_{jk}\,H$. The question is then how does one go *off-diagonal* when describing intermediate quantum states...Using Dirac's *contact form* for quantum scattering, the momentum of the intermediate states is conserved, thereby requiring from the algebra that when the energy is off shell, the Lorentz frame velocities are different for intermediate states, $\underline{v} = \dfrac{\underline{P}}{E} \neq \dfrac{\underline{P}}{E'} = \underline{v}'$. Having differing frames of reference used to describe a given scattering process is geometrically counter-intuitive. However, if one utilizes Dirac's *point form*[14,15] for quantum dynamics, the Lorentz frame is conserved $\dfrac{\underline{P}}{E} = \underline{v} = \underline{v}' = \dfrac{\underline{P}'}{E'}$.
- The spectating cluster kinematics enters the equations only parametrically.
- The kinematic description becomes quite complicated unless the right geometric parameters are chosen to depict the quantum states. The natural parameters needed to describe the dynamics are the invariant rest energy of the system as a whole, the velocities of the overall system and dynamical cluster, and the angular orientation (or alternatively, the angular momentum) of the dynamical cluster $M, \underline{u}, \underline{u}_a, \hat{q}_a$.

ully Lorentz covariant, and when *on-shell* generates full four-momentum conservation (forcing the external legs of the system to be *on-diagonal*). However, the kinematics between external and intermediate quantum states (off-diagonal dynamics) has been chosen to insure Lorentz frame (3-velocity) conservation (Dirac point form) rather than 3-momentum conservation (contact form). Unitarity (probability conservation) and cluster decomposability is assured. The complex analytic extension of the invariant energy of a given cluster (the *off-shell* behavior) only parametrically affects the kinematics of the other clusters, i.e., the internal dynamics of one cluster does not alter the energy spectrum of another. Finally, all amplitudes have well defined non-relativistic limits consistent with known phenomenology.To examine the subtleties of quantum disentangled events, Figure 8 demonstrates a diagrammatic representation of the Compton scattering of a photon from a charged particle, including all systems concerned with the scattering.

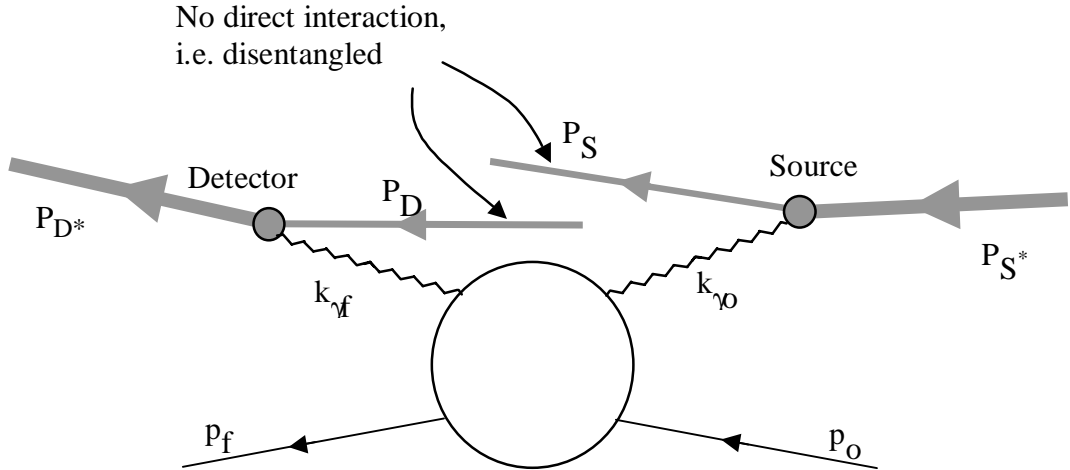

**FIGURE 8.** Whole universe view of Compton scattering.

This formulation is consistent the Wheeler-Feynman treatment of photons[16]. In the figure, the source of the incident photon and the detector of the scattered photon have no direct entanglement. The source and detector are only related through the scattering photons and charged particle. In this sense, the source and the detector are classical systems relative to each other.

To summarize, the dis-entanglement of relativistic dynamic quantum clusters that is necessary for correspondence with classical dynamics requires that the geometric aspects of the kinematics associated with the Lorentz/Poincare transformation properties of a given cluster must be separate and distinct from the internal coherent descriptions and off-shell analytic behaviors of disparate clusters. The solution requires the proper cluster independence of the geometric parameters describing the kinematics between subsystems from the internal quantum dynamics associated with the description of an interacting system in terms of the boundary (i.e., only self-interacting) states.

## Macroscopic Quantum Co-Gravitation

This section will develop constituent quantum scalars that satisfy expected physical properties, and can be combined to consistently co-generate the background geometry. This will be accomplished by first constructing a Lagrangian for general gravitating quantum scalars that are dimensionally consistent with the geometry. The form of the Lagrangian should incorporate additive energy-momentum contributions related to the phase coherence satisfying Overhauser's experiment, allowing algebraic solutions to determine the relative density fractions needed to satisfy Einstein's equation.

Utilizing intuitions derived from the flat space-time solution, the present treatment will use an affine flow[17] (or proper time[18]) parameterization in developing Lagrangians with substantive/material flows to incorporate the Principle of Equivalence. These quanta will gravitate in the background described by the metric (4), which gives an Einstein tensor of the form

$$((G^{\mu\nu})) = \begin{pmatrix} 0 & 0 & 0 & 0 \\ 0 & -\dfrac{\dot{R}_M}{r^2\sqrt{\zeta}} & 0 & 0 \\ 0 & 0 & -\dfrac{\dot{R}_M}{4r^4\sqrt{\zeta}} & 0 \\ 0 & 0 & 0 & -\csc^2\theta\dfrac{\dot{R}_M}{4r^4\sqrt{\zeta}} \end{pmatrix}, \qquad (10)$$

or in the more useful mixed contravariant-covariant form

$$((G^{\mu}{}_{\nu})) = \begin{pmatrix} 0 & 0 & 0 & 0 \\ -\dfrac{\dot{R}_M}{r^2} & -\dfrac{\dot{R}_M}{r^2\sqrt{\zeta}} & 0 & 0 \\ 0 & 0 & -\dfrac{\dot{R}_M}{4r^2\sqrt{\zeta}} & 0 \\ 0 & 0 & 0 & -\dfrac{\dot{R}_M}{4r^2\sqrt{\zeta}} \end{pmatrix}. \qquad (11)$$

The form is seen to have a vanishing total energy density component.

### *Gravitating Scalar Field*

The gravitating massive (or massless) scalar quanta will have a Lagrangian linear in its affine derivative, given by

$$\boldsymbol{L}_s = -\sqrt{-g}\left[\dfrac{i\hbar c}{2}u^{\beta}\left(\psi_s^*\partial_{\beta}\psi_s - \partial_{\beta}\psi_s^*\,\psi_s\right) - m_s c^2 \psi_s^*\psi_s\right]. \qquad (12)$$

The field will be assumed to be described by a complex scalar of the form

$$\psi_s = |\psi_s|\,e^{i\xi_s}. \qquad (13)$$

The Euler-Lagrange equations for independent variations of the components of the complex field result in a probability conservation condition

$$\dfrac{1}{\sqrt{-g}}\partial_{\beta}\left(\sqrt{-g}\,|\psi_s|^2\,u^{\beta}\right) = 0, \qquad (14)$$

and a phase coherence condition on the field

$$u^\beta \partial_\beta \xi_s = -\frac{m_s c}{\hbar}. \tag{15}$$

The substituted extremal form of the Lagrangian vanishes, as expected for these pressureless, non-interacting quanta.

$$L_s[\psi_{extremal}, \partial_\mu \psi_{extremal}] = 0. \tag{16}$$

Using standard techniques in quantum field theory, the energy-momentum tensor of the field can be developed:

$$T_s{}^\beta{}_\mu = \frac{\partial L_s}{\partial(\partial_\beta \Psi_s)} \partial_\mu \Psi_s - \delta^\beta_\mu L_s = \sqrt{-g}\,\hbar c u^\beta (\partial_\mu \xi_s) |\psi_s|^2. \tag{17}$$

By direct dimensional analysis, this energy-momentum density has the correct functional dependency needed to contribute to the Einstein tensor Eq. (11). In particular, it should be noted that there is a linear connection of the phase dynamics to the energy-momentum tensor. This means that the energy and momentum associated with the phase of the coherent "wavefunction" can linearly contribute to the overall local energy-momentum tensor that is connected to the geometry through Einstein's equation.

The phase coherence relation Eq. (15) can be further examined for consistent solutions. A generic functional form for the phases is given by

$$\partial_{ct} \xi_s = \frac{E^*_m}{\hbar c}(u_{ct} + Q u^r),$$
$$\partial_r \xi_s = \frac{E^*_m}{\hbar c}(u_r - Q u^{ct}), \tag{18}$$
$$u^\beta \partial_\beta Q = \partial_{ct} u_r - \partial_r u_{ct} - (\partial_\beta u^\beta) Q.$$

The third equation for the general parameter $Q$ results from an integrability requirement on the phases $\xi_s$.

Solutions to Eq. (14) represent the trajectories of a gravitating scalar field of this type. The magnitude squared of the scalar field represents the probability density of the field when properly normalized. This quantity is represented in Figure 9.

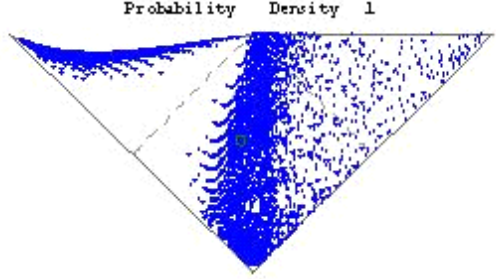

**FIGURE 9.** Probability density of a gravitating massive scalar field.

In the diagram, the computer is instructed to place a blue pixel at space-time location labeled *(ct,r)* if the value of the probability density is greater than its value at a normalization point, in this case taken on the radial mass scale. The computer is also instructed to place a red pixel at that location if the value of the density being plotted is less than the negative of the value at the normalization point. The numerical solution is considered to be non-physical if the probability density is not positive semi-definite.

The overall strategy has been to develop cluster decomposable co-gravitating fields that can combine to generate the given background space-time metric. The linear relationship of the energy and momentum form in the *phase* of the field to the energy-momentum tensor allows one to construct superposable contributors to the Einstein tensor describing the geometry. The remaining contribution to the energy-momentum of the geometry should be due to a core gravitating field directly associated with the collective system of co-gravitating quanta. One can then develop this (expectedly real) core gravitating field to incorporate the overall symmetries of the system (as occurs, for instance, in utilizing Gauss' Law in electromagnetism). Once a functional form for the core gravitating field has been established, the form of the constituent contributions from the radiating gravitating scalars to the overall energy-momentum tensor in Einstein's equation can be algebraically solved for self-consistency:

$$G^{\beta}{}_{\mu} = -8\pi \frac{L_P^2}{\hbar c}\left(\sum_s T_s{}^{\beta}{}_{\mu} + T_{core}{}^{\beta}{}_{\mu}\right). \qquad (19)$$

The task of constructing a core gravitating field consistent with the co-gravitating scalars is next undertaken. As previously stated, the core gravitating field is expected to be real, should incorporate the symmetries of overall geometry, and is likewise expected to co-gravitate with the other constituents of the energy-momentum of the overall system. For this presentation, the core gravitating field will be constructed to involve massless core field quanta. A Lagrangian form that satisfies these criteria is given by

$$L_{core} = -\sqrt{-g}\,\hbar c\left[\psi_{c+}\left(u^\beta \partial_\beta \psi_{c+} + \frac{1}{2}u^\beta \partial_\beta\left(\log\sqrt{-g}\right)\psi_{c+}\right) + \right.$$
$$\left. -\psi_{c-}\left(u^\beta \partial_\beta \psi_{c-} + \frac{1}{2}u^\beta \partial_\beta\left(\log\sqrt{-g}\right)\psi_{c-}\right)\right]. \quad (20)$$

(If the reader is interested in the possibility of a stationary core field, only a slight modification of this Lagrangian form is required.) The resultant Euler-Lagrange equation for this real field is given by

$$\sqrt{-g}\,\psi_{c\pm}\partial_\beta u^\beta = 0. \quad (21)$$

This equation is satisfied by *any* massless field in the given space-time background. As was the case for the radiating gravitating quanta, the contribution of the core gravitating field to the overall energy-momentum tensor satisfies all dimensional requirements needed to connect with the Einstein tensor of the geometry. The connection in the Einstein equation for isotropic fields then constrains the form of the core field via the relation

$$-\frac{\dot R_M}{4r^2\sqrt{\zeta}} = G^\theta{}_\theta = -8\pi\frac{L_P^2}{\hbar c}T_{core}{}^\theta{}_\theta = 8\pi\frac{L_P^2}{\hbar c}L_{core}/\sqrt{-g}. \quad (22)$$

Equations (20) and (22) then define the core gravitating field consistent with the assumption of a spherically symmetric distribution of the radiating scalars.

The numerical solution for the core gravitating field satisfying Eq. (22) is shown in Figure 10.

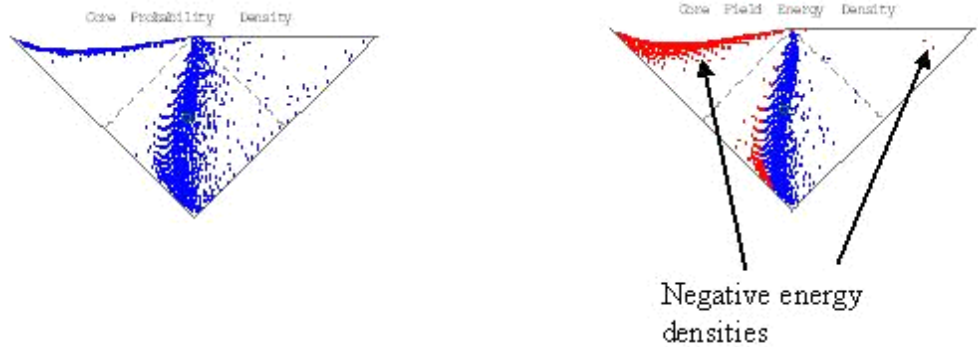

**FIGURE 10.** Core gravitating field probability density and energy density.

Again, the computer was instructed to place a blue pixel at *(ct,r)* if the value of the plotted density was greater than the absolute value of the density at the normalization point indicated by the center of the small circle near the radial mass scale, and a red pixel if the density was less than the negative of the normalizing value. The probability density is everywhere positive semi-definite, as should be the case for physical fields. However, the energy density of the core gravitating field is seen to have negative values near and within the horizon, as well as far from the horizon. In particular, since the overall energy density of the space-time vanishes from Eq. (10), the negative values of the core field density far from the horizon are physically necessary if the contributions of the radiating gravitating quanta in the asymptotic region are to take the conventional positive forms.

Given the numerical solutions for the core gravitating field, co-gravitating radiating scalar fields must algebraically satisfy Einstein's equation. Since the overall energy-momentum tensor consists of a sum of otherwise linearly independent contributors, such solutions are always possible. It only remains to determine whether such solutions can satisfy the physical boundary conditions of a given system. In order to demonstrate that this can be done, a particular solution will be constructed. This minimal solution notes that the mixed form of Einstein's tensor in Eq. (11) has only 4 independent components that remain to be satisfied after the form of the core gravitating field resulting from Eq. (22) has been constructed. Therefore, a particular solution involving 4 coherent quantum types has been found. Each of the four types $j=1 \rightarrow 4$ will be assumed to have $N_j$ quanta present, and the number densities consistent with Einstein's equation will be determined. The coherently radiating quanta then satisfy

$$G^\beta{}_\mu = -8\pi \frac{L_P^2}{\hbar c} \left( \sum_s T_s{}^\beta{}_\mu + T_{core}{}^\beta{}_\mu \right),$$

$$\sum_s T_s{}^\beta{}_\mu = \sum_{j=1}^4 N_j u_j^\beta (\partial_\mu \xi_j) |\psi_j|^2.$$

(23)

Dimensional analysis allows the calculated number densities to be expressed in the form

$$\mathbf{N}_j(\zeta) \equiv \frac{E_m}{\hbar c} L_P^2 r^2 N_j(\zeta) |\psi_j(ct,r)|^2.$$

(24)

The coherent emission/absorption of $N_j$ quanta of type j by the black hole then co-generates the geometry of the black hole. Each quantum satisfies local probability conservation and phase coherence consistent with the experimentally observed behaviors of gravitating quanta.

It is gratifying that there are many solutions that generate the given geometry. The general boundary conditions required for a complete solution are particle masses, particle boundary (radial) velocities and directions (outgoing, ingoing, or stationary), and the phase information of each particle on the boundary. A particular solution is shown in Figure 11.

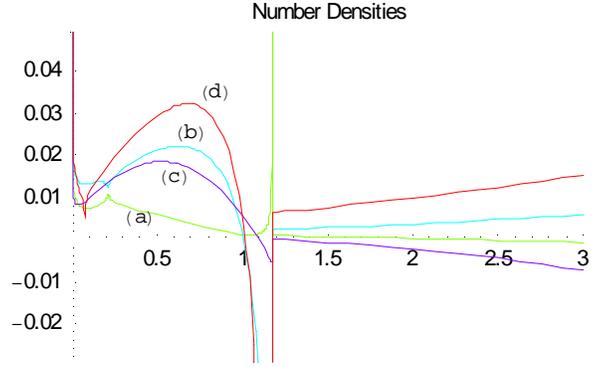

**FIGURE 11.** Dimensionless number densities of co-gravitating quanta vs. $\zeta$.

For this solution, the four quantum types were taken to be two massive stationary quanta with linearly independent phase boundary conditions (solutions a and b), an outgoing massless quantum with positive $Q$ phase boundary condition (from Eq. (18), solution c), and an ingoing massless quantum with negative $Q$ phase boundary condition (solution d).

Since this presentation is not concerned with the microscopic behaviors of the transition from the excreting black hole geometry to the low curvature final state, the co-gravitation of the quanta will only be examined in the region of the space time for which the dynamic metric form Eq. (4) is valid. The solution for the dimensionless number density $N_{(a)}(\zeta)$ from Eq. (24) for the stationary massive particles satisfying the solution type (a) in Figure 11 is therefore plotted as a density on the lower portion of the Penrose diagram for the black hole in Figure 12.

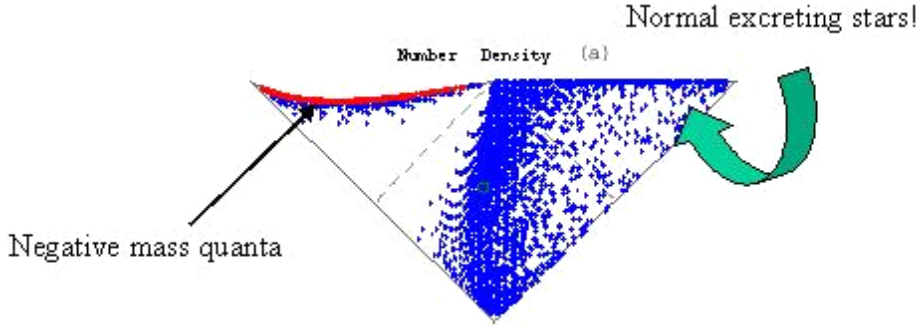

**FIGURE 12.** Penrose density plot of dimensionless number density of stationary radiations of type (a).

It is important to note that the dimensionless density here plotted is proportional to the inverse Compton wavelength of the particle multiplied by the number density. In regions far from the horizon, this quantity is always positive, which is the exterior region relevant for normal excreting stars. However, since number density must be positive semi-definite, this means that in the interior region of an excreting black hole solution, there must be some negative mass quanta. This seems to be an artifact of the behavior of the 4-velocities of particles as they traverse the radial mass scale $R_M$. The radial component of the four-velocity often changes sign at this scale, as can be seen from the form of the metric .(4) Similar plots can be developed for each of the particle types in Figure 11.

# GLOBAL STRUCTURE OF A FLUID COSMOLOGY

The primary motivation for the author's interest in the behavior of the dynamic black hole previously discussed was his interest in any dynamic horizon that might be associated with cosmological dark energy[19,20]. The insights gained by examining quantum behaviors on a spatially coherent spherically symmetric space-time consistent with familiar Newtonian gravitation in the weak-field, slow-moving limit should considerably aid in understanding analogous behaviors for the cosmology as a whole. The presentation will therefore end with a discussion of the global structure of a multi-fluid cosmology consistent with observed phenomenology.

There are several observed parameters of the universe that are of interest to this discussion. One should note that the Friedmann-Lemaitre equations, which are fluid dynamic equations that describe the evolution of the energy density of the universe, are spatially scale invariant, but *not* temporally scale invariant. There is apparently a beginning time $t_o \approx 13.7(\pm 0.2)$ billion years ago which represents the earliest backwards-looking extrapolation of the standard model expansion called the *big bang*. The physics during these earliest moments is yet to be understood, since the energies involved exceed known high energy phenomenology. The observation of standard candles, galactic rotation curves, features of the cosmic microwave background, and other phenomena, have been incorporated into the standard cosmological model[21] that exploits the relative densities for photons, baryons, dark matter, and dark energy given by $\Omega_\gamma \sim 4.9 \times 10^{-5}$, $\Omega_b \sim 0.04$, $\Omega_{dm} \sim 0.22$, and $\Omega_\Lambda \sim 0.73$ observed today. The measured values are consistent with a cosmology consisting of a thermal fluid with remnant pressureless matter and dark energy on a spatially flat space-time background. These fluids will be incorporated in the non-singular cosmology that follows.

## Fluid Cosmology

Consider the general form of the space-time generated by an ideal fluid,

$$T_{\mu\nu} = P g_{\mu\nu} + (\rho + P) u_\mu u_\nu, \qquad (25)$$

with consistency condition

$$u_\mu g^{\mu\nu} u_\nu = -1. \qquad (26)$$

The geometry will be assumed to have isotropic flows, $u_\vartheta = 0 = u_\varphi$. Einstein's equation can then be used to determine the fluid parameters associated with a fluid consistent geometry:

$$P = T^\vartheta_\vartheta = T^\varphi_\varphi = -\frac{c^4}{8\pi G_N} G^\vartheta_\vartheta,$$

$$\rho = 3P + \frac{c^4}{8\pi G_N} G^\mu_\mu,$$

$$u_0^2 = \frac{(T_{00} - g_{00}P)}{(\rho + P)} = -\frac{\frac{c^4}{8\pi G_N}G_{00} + g_{00}P}{(\rho + P)}, \quad (27)$$

$$u_r^2 = \frac{(T_{rr} - g_{rr}P)}{(\rho + P)} = -\frac{\frac{c^4}{8\pi G_N}G_{rr} + g_{rr}P}{(\rho + P)}.$$

Working the problem in reverse, given any arbitrary metric form, fluid parameters can be defined using Eq. (27). The fluid can be physical if the consistency condition on the 4-velocities from Eq. (26) is satisfied.

A dynamic metric analogous to Eq. (4) for a fluid cosmology can be constructed. One desires a metric that incorporates cosmological scales which evolve with differing temporal dependencies. What has been developed are non-orthogonal coordinates that mix a scale factor that would become like de Sitter space if diagonalized using a change in temporal coordinate with one that would become like Robertson-Walker space if diagonalized using a change in the radial coordinate. The dynamic scaled fluid cosmology has a metric of the form[22]

$$ds^2 = -\left(1 - \frac{r^2}{R_\rho^2(ct)}\right)(dct)^2 - \frac{2r}{R_\rho(ct)}dct\,dr + \left[dr^2 + r^2\left(d\theta^2 + \sin^2\theta\,d\varphi^2\right)\right]. \quad (28)$$

This metric form introduces a microscopic *fluid scale* $R_\rho$ that is directly related to the usual Robertson-Walker scale $a$ (within a diagonal metric form that shares the same temporal coordinate) via

$$\frac{1}{R_\rho^2} \equiv \frac{8\pi G_N}{3c^4}\rho \quad, \quad \frac{\dot{a}}{a} = \frac{1}{R_\rho} \quad, \quad a(ct) = R_\rho(0)\exp\left(\int_0^{ct}\frac{dct'}{R_\rho(ct')}\right). \quad (29)$$

The dynamic fluid metric form given by Eq. (28) has several features of interest. This metric incorporates a fluid scale that can evolve away from/towards a cosmological pseudo-constant in Einstein's equation. Therefore, an early inflationary period can evolve into a final dark energy dominated epoch through an intermediate big bang cosmology in a straightforward manner. The metric allows the dynamic evolution of dark energy and quantum evolution of cosmology without a need to introduce any true cosmological constants. It provides a very convenient form for studying the early cosmology, and as will be seen, need not have primordial singular behavior.

## Multi-Fluid Cosmology

The Einstein equation associated with the metric form (28) generates a fluid dynamic cosmology satisfying

$$\frac{d\rho}{dct} = -\sqrt{\frac{24\pi G_N \rho}{c^4}}(\rho + P). \tag{30}$$

This equation can be seen to be equivalent to the Friedmann-Lemaitre equations of standard cosmology, only written completely in terms of the physical densities. A multi-fluid cosmology can next be developed by decomposing the physical densities into components due to dark/primordial energy, radiation, and dust:

$$\rho = \rho_{DarkEnergy} + \rho_{radiation} + \rho_{dust} \quad , \quad \rho_{DarkEnergy} = \rho_{primordial} + \rho_{remnant}. \tag{31}$$

The dynamical equation (30) can likewise be decomposed into coupled rate equations.

$$\left(\frac{d\rho_{DarkEnergy}}{dct}\right) = -D_{DE \to rad}(ct) - \frac{3}{R_\rho}\left(\rho_{DarkEnergy} + P_{DarkEnergy}\right). \tag{32}$$

$$\left(\frac{d\rho_{radiation}}{dct}\right) = +D_{DE \to rad}(ct) - \frac{4}{R_\rho}\rho_{radiation} - \Theta(\rho_{radiation} - \rho_{threshold})\frac{\rho_{radiation}}{c\tau_{rad \to dust}}. \tag{33}$$

$$\left(\frac{d\rho_{dust}}{dct}\right) = +\Theta(\rho_{radiation} - \rho_{threshold})\frac{\rho_{radiation}}{c\tau_{rad \to dust}} - \frac{3}{R_\rho}\rho_{dust}. \tag{34}$$

The second and third equations have the expected equations of state for radiation and pressureless dust directly incorporated. The chosen form presumes that the primordial energy dissolves into radiation. That radiation then produces remnant dust due to asymmetries in microscopic interactions above a threshold density. The incorporation of the microphysics can be modified in a straightforward manner if necessary, as long as the overall conservation properties are respected. Numerical solutions of these equations can directly be obtained once the microscopic behavior of the dissolution of the primordial energy density into radiation, $D_{DE \to rad}$, is incorporated.

## A Cosmology with Remnant Dark Energy

For the numerical calculation presented, the dark energy density was given a Gaussian form

$$\rho_{DarkEnergy} = \rho_\Lambda + \left(\rho_{primordial} - \rho_\Lambda\right)e^{(t/\tau_{DE \to rad})^2}. \tag{35}$$

where $\rho_\Lambda$ is the remnant dark energy of the type measured today, $\rho_{primordial}$ is the (initial) primordial energy density, and $\tau_{DE \to rad}$ is the microscopic time scale associated with the dissolution of primordial energy into radiation. For convenience, it was assumed that the dark energy satisfied the equation of state $P_{DE}=-\rho_{DE}$, consistent with the observed equation of state for the remnant dark energy. It should be noted that the primordial dark energy need not generally satisfy the same equation of state satisfied by the remnant dark energy. Also, for clarity of presentation, the actual expected temporal and density scales of standard cosmology have only been qualitatively incorporated. The coupled rate equations (32), (33), and (34) can then be numerically solved. For the parameters chosen for this presentation, the temporal dependencies of fluid densities near the beginning of radiation domination are demonstrated in Figure 13.

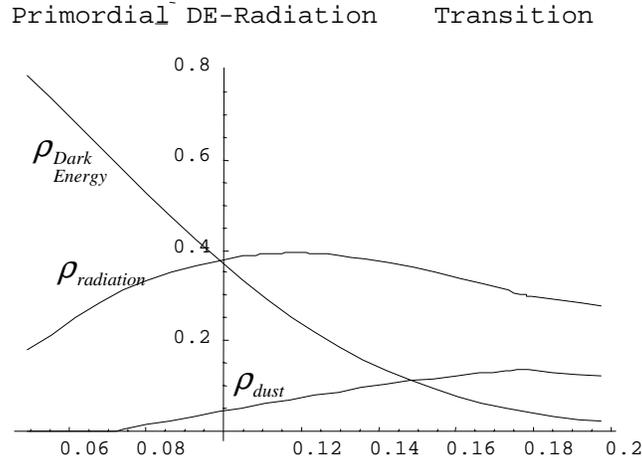

**FIGURE 13.** Numerical form of early cosmology fluid components.

The parameters have been chosen such all initial energy density is primordial dark energy which undergoes a dissolution into radiation that dominates an early hot thermal cosmology and a small remnant final state dark energy that dominates the cosmology at very times. Sufficiently dense radiation generates the remnant dust which during later intermediate times dominates the other fluid components.

Once the physical densities are known, the geometrical scales are likewise determined, and the Penrose diagram for the geometry can be constructed. This diagram, which displays the global causal structure of the space-time, is displayed in Figure 14.

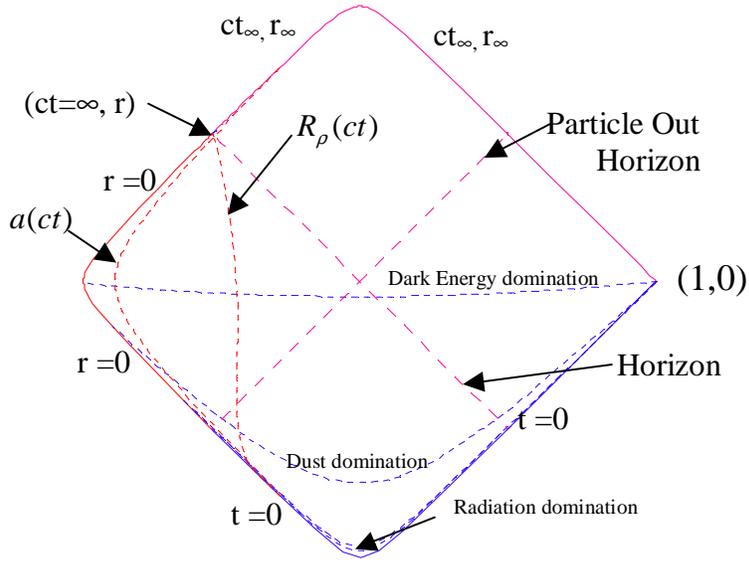

**FIGURE 14.** Features of the Penrose diagram for a non-singular multi-fluid cosmology.

This diagram has several features of interest. Foremost is that it consists of only 3 boundaries, an initial space-like volume, a final space-like future infinity, and a time-like *center*. The conformal coordinates used to construct the Penrose diagram were chosen to vanish at the point of intersection of the deSitter-like horizon resulting from the non-vanishing remnant dark energy and the particle out horizon, which is the most distant light-like surface that can ever receive a communication from the center. The center is, of course, not special, due to the cosmological principle, which is demonstrably apparent in the diagonalized Robertson-Walker form of this cosmology. The dynamic fluid scale $R_\rho$ is seen to coincide with the horizon at a unique future infinity point on the diagram. The Robertson-Walker scale has been chosen to initially coincide with the (initial, primordial) fluid scale, making the geometry non-singular. It should be noted that there is only one point on the diagram, labeled (1,0), for which any conformal coordinate coincides to an extremal value. Space-like volumes corresponding to the times of the beginning of radiation domination, dust domination, and remnant dark energy domination, are clearly labeled on the diagram.

A grid of fixed coordinate curves can be placed on the Penrose diagram in order to map points in the space-time. From the form of the metric Eq. (4), surfaces of fixed radial coordinate r are seen to be spheres of fixed static areas $4\pi r^2$. Coordinate grids for the dynamic fluid coordinates *(ct,r)* and the corresponding Robertson-Walker coordinates $\left(ct, r_{RW} = \dfrac{r}{a(ct)}\right)$ are displayed in Figure 14.

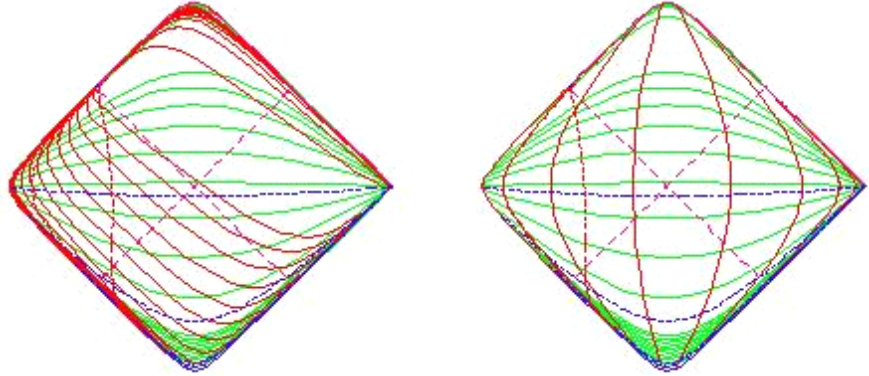

**FIGURE 15.** Static area and Robertson-Walker coordinates.

The diagram on the left represents the *(ct,r)* static area coordinates, while that on the right represents the Robertson-Walker coordinates. Curves of fixed r (transverse area) are all time-like to the left of the fluid scale, which represents a dynamic surface where ingoing light-like trajectories are momentarily stationary in the radial coordinate. These fixed radial curves are initially graded in tenths, then units, and decades of the given scale. Fixed area curves all initiate at various points on the volume *t=0* and terminate at the previously mentioned future infinity point shared by the fluid scale and the horizon. A close inspection of the initial behavior of the fluid scale demonstrates that it is only very slowly changing in radial coordinate initially, as expected for an early-stage inflation. The curves of fixed temporal coordinates, initially graded in tenths, units, then decades of the given scale, all initiate on the time-like surface *r=0* and terminate at the extremal point (1,0). These fixed time curves remain space-like volumes throughout the cosmology.

The diagram on the right has the same fixed temporal volumes as that on the left. However, the curves of the fixed Robertson-Walker radial coordinate $r_{RW}$ remain time-like surfaces throughout the space-time. Each of these surfaces, graded in units of the given scale, initiate on the *t=0* surface and terminate on the static infinite area surface $r=\infty$. Curves of fixed Robertson-Walker radial coordinate represent inertial co-moving centers in the cosmology.

The existence of the finite scaled horizons directly exhibits regions of the space-time that are causally inaccessible throughout all time. Events that occur in the far right quadrant of the space-time can never have any communicative effect on an observer at the center r=0. Similarly, events in the far left quadrant can have no effect on any observer whose initial location was to the right of the horizon in the far right quadrant. However, one should note that there can be space-like coherent events across either of these causally disjoint regions of the space-time. This parameterization is very convenient for examining quantum cosmological behaviors, which is work that is presently underway.

It should be straightforward to directly incorporate the standard cosmological scales into this formulation. If actual values had been utilized in this presentation, the boundaries of the Penrose diagrams would have appeared indistinguishable from light-like surfaces, and the early transition features would not have been distinguishable from the volume t=0. However, the presenter has done an analysis of the evolution of the scales of standard cosmology elsewhere[20, 23]. The evolution of the cosmology during the period for which the primordial energy density is negligible can be modeled using the Friedmann-Lemaitre equations of standard cosmology. Given a particular Robertson-Walker scale at some fixed time, that particular scale can be extrapolated back to a time that it expands at the speed of light c. From that earliest time considered, there was a period of deceleration through radiation and dust domination, followed by acceleration towards a de Sitter-like expansion associated with the fixed cosmological constant of standard cosmology. The observed remnant dark energy scale is connected to the cosmological constant via the relation $R_\Lambda \equiv \sqrt{\frac{3}{\Lambda}} \approx 10^{28} cm \approx 1.6 \times 10^{10} ly$. Using this observation, the redshift $z$ relative to present observations $1 + z \equiv \frac{a_o}{a(ct)}$ and the rate of scale expansion relative to the speed of light c are plotted in Figure 16.

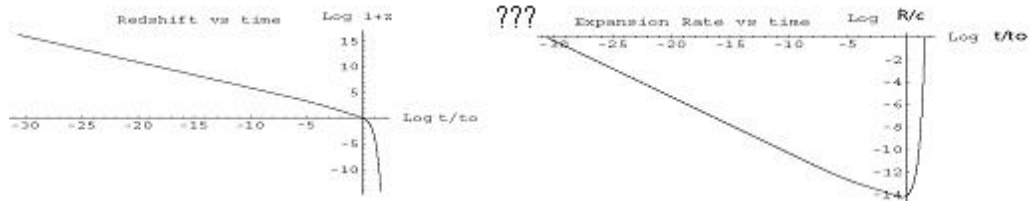

**FIGURE 16.** Log-log plots of the cosmological redshift (left) and scale expansion rate.

For both plots, the present time corresponds to the origins. In particular, the plot on the right demonstrates that for standard cosmology, there are two times for which a given scale expands at a rate equal to the speed of light. For the dynamic multi-fluid model developed in this presentation, the rate of the expansion of the Robertson-Walker scale is demonstrated in Figure 17.

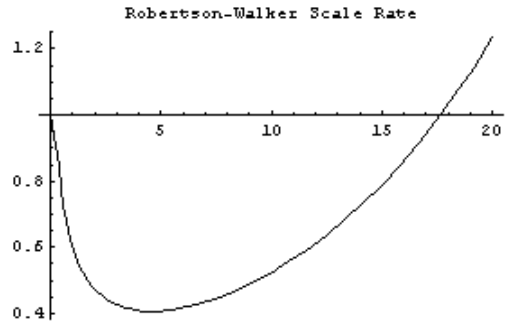

**FIGURE 17.** Expansion rate of Robertson-Walker scale in multi-fluid cosmology.

Although this is not a log-log plot, it is seen to have the expected form. Furthermore,

the behavior at the earliest times can be directly studied, which was the primary motivation for the development of this dynamic multi-fluid model.

## CONCLUSIONS AND DISCUSSION

Due to quantum measurability constraints, dynamic geometries have been shown to provide excellent laboratories for exploring co-gravitating quantum behaviors. Insights into developing consistent models come from experimental results on gravitating coherent systems, as well as the fundamental quantum mechanics of relativistic systems in flat space-time. In particular, the development of cluster decomposable quantum forms has been shown to provide a framework for algebraic solutions to Einstein's equation for co-gravitating quanta. A consistent, generic system of co-gravitating radiating scalars and a massless core field describing an excreting spatially coherent black hole has been presented.

The insights gained by modeling a dynamic black hole can be applied to the universe as a whole. A geometry with early spatial coherence directly addresses the horizon problem as usually put forth in standard cosmology, and need not exhibit a primordial singularity. For these reasons, a model that describes a multi-fluid cosmology consisting of primordial and remnant dark energies, radiation, and dust, has been developed. The Penrose diagram of the global causal structure of this cosmology directly exhibits its spatial coherence and horizon structure. The presented model is qualitatively consistent with standard cosmology. The presenter is presently exploring quantum behaviors on this multi-fluid cosmology, with a goal of developing a quantum cosmology consistent with observed phenomenology.

## ACKNOWLEDGMENTS

The presenter gratefully acknowledges useful discussions with the following: Stephon Alexander, Marcus Alfred, James Bjorken, Beth Brown, Tehani Finch, Larry Gladney, Keith Jackson, E.D. Jones, Petero Kwizera, Alex Markevich, Harry Morrison, H. Pierre Noyes, Michael Peskin, Paul Sheldon, and Lenny Susskind.